\documentclass[aps,physrev,preprint,superscriptaddress]{revtex4-2}

\usepackage{physics, comment}
\usepackage{subcaption}
\usepackage{graphicx}
\usepackage{appendix}
\usepackage{amsmath}
\graphicspath{{figures/}{others/}}

\usepackage[colorlinks=true,linkcolor=red,urlcolor=blue,citecolor=blue]{hyperref}

\begin{document}
\title{Impact of Correlated Test-Mass Noise on Stochastic Gravitational-Wave Background Searches with Space-Based Interferometers}

\author{Jing-yi Wu}
\affiliation{School of Astronomy and Space Science, University of
	Chinese Academy of Sciences (UCAS), Beijing 100049, China}

\author{Yong Tang}
\affiliation{School of Astronomy and Space Science, University of
	Chinese Academy of Sciences (UCAS), Beijing 100049, China}
\affiliation{School of Fundamental Physics and Mathematical Sciences,
	Hangzhou Institute for Advanced Study, UCAS, Hangzhou 310024, China}
\affiliation{International Center for Theoretical Physics Asia-Pacific, Beijing 100190, China}

\begin{abstract}
The stochastic gravitational-wave background (SGWB) is a major science target of future space-based interferometers because it encodes information about unresolved astrophysical populations and physical processes in the early Universe. Its stochastic nature, however, means that its detection and parameter inference rely on distinguishing its statistical contribution to the data from that of instrumental noise. In particular, both the SGWB and instrumental noise contribute to the data covariance, so an incomplete noise model can be misinterpreted as part of the SGWB and thereby bias the inferred spectrum and its parameters.
Here we investigate this effect in the presence of correlations between the acceleration (ACC) noises of adjacent test masses (TMs) in each spacecraft. {We simulate the correlated noise at the single-link level, propagate it through the moving-orbit response into the time-delay interferometry (TDI) observables}, and analyze the resulting TDI auto spectra with two Bayesian frequency-domain templates: the Corr. Template includes the correlated-noise covariance, whereas the Non-Corr. Template neglects it. 
We find that, when correlated ACC noise is present in the mock data, the Non-Corr. Template exhibits systematic parameter offsets that generally become more pronounced with increasing correlation amplitude. In contrast, the Corr. Template recovers the injected SGWB and instrumental-noise parameters substantially more accurately. These results demonstrate that correlations in test-mass ACC noise can constitute a relevant systematic for space-based SGWB searches and motivate their explicit inclusion in mission-realistic noise models and inference pipelines.
\end{abstract}

\maketitle
\newpage

\section{Introduction}
Space-based laser interferometers such as LISA \cite{2017arXiv170200786A}, Taiji \cite{10.1093/nsr/nwx116}, and TianQin \cite{Luo_2016} will open the millihertz gravitational-wave band. This frequency range provides access to SGWB contributions from both astrophysical populations \cite{phinney2001practicaltheoremgravitationalwave,Regimbau_2006,Regimbau_2007,Marassi_2011,Tania_Regimbau_2011,Zhu_2013,Fan_2018,Babak_2023,Bonetti_2020,PhysRevD.108.103039} and cosmological processes in the early Universe \cite{Starobinsky:1979ty,Ananda_2007,Easther_2006,Easther_2007,Garc_a_Bellido_2007,Dufaux_2007,Garc_a_Bellido_2008,Bethke_2013,Bethke_2014,Kamionkowski_1994,Apreda_2002,Grojean_2007,Caprini_2008,Huber_2008,Caprini_2009,Leitao_2012,Hindmarsh_2014,Giblin_2014,Jinno_2017,Caprini_2016,Chao_2017,Weir_2018,Hindmarsh_2019,PhysRevD.31.3052,PhysRevD.42.354,PhysRevD.45.3447,Caldwell_1996,Figueroa_2013,Blanco_Pillado_2017,Liu_2021,Caprini_2018,Auclair_2023}. The diversity of these sources makes the SGWB scientifically valuable, but it also makes accurate spectral reconstruction essential for interpreting a future detection.

Unlike resolved compact-binary signals, an SGWB does not produce an identifiable waveform that can be singled out in the time series. Instead, it manifests itself statistically in the detector output, typically as additional power distributed over a range of frequencies. Instrumental noise can produce a similar excess over the same frequency range, so SGWB inference requires separating the gravitational-wave contribution from that of the instrument. This separation problem has motivated a broad range of parametric and semi-parametric inference methods \cite{Adams_2010,Adams_2014,Caprini_2019,Flauger_2021,Baghi_2023,PhysRevD.109.042001,Pozzoli_2024,csx9-9trp,PhysRevD.107.123531,Wang_2022,Braglia_2024,Caprini_2024,Blanco-Pillado_2025,Gammal_2025}, as well as studies incorporating realistic complications such as unequal and time-dependent arm lengths, unequal or nonstationary noise, limitations of null-channel noise monitoring and power spectral density (PSD) estimation, and the time dependence, anisotropy, and multiple components of astrophysical foregrounds \cite{PhysRevD.105.062006,PhysRevD.107.082004,Wang_2024,Kume_2025,dcb6-1jsl,Digman_2022,Ren_2023,Du:2025xdq,Li_2025,gpmh-1hqx,Banagiri_2021,Rieck_2023,TianKouDing_2024,BloomCriswellMandic_2024,Digman_2023,criswell2025flexiblespectralseparationmultiple}. These studies emphasize that reliable SGWB inference depends on an adequate description of both the signal and instrumental noise, since any relevant contribution that is omitted or inaccurately modeled can be absorbed into the inferred SGWB spectrum and thereby bias its recovered parameters.

Evidence from LISA Pathfinder makes correlated ACC noise a concrete instrumental concern: environmental and control variables can exhibit coherence with differential acceleration \cite{Boileau_2022}. Although this observation does not by itself establish the level or structure of correlations between adjacent TMs expected in a future mission, it demonstrates that ACC noise can retain signatures of shared disturbances. In a future space interferometer, adjacent TMs may similarly respond to common subsystems like magnetic, thermal, residual-gas, and control-related disturbances \cite{2017arXiv170200786A,PhysRevLett.120.061101}. Correlated components of this noise propagate through the interferometer response into the TDI observables and therefore contribute to the same spectral structure from which the SGWB is inferred. If such correlations are present in the data but omitted from the inference model, the resulting mismatch can be absorbed by the parameters retained in the template, leading to biased inference of both the SGWB and instrumental noise.

In this work, we adopt a forward-modeling framework to quantify this effect. We simulate mock data containing ACC noise, optical metrology system (OMS) noise, SGWB, and two phenomenological cross spectra describing correlations between adjacent TMs: the white-noise (WN) spectrum and the PSD-scaled (PS) one. The link-level auto- and cross-spectral contributions are then propagated through the moving-orbit response into the TDI observables. We analyze the $A$ and $E$ channels' auto spectra using two Bayesian frequency-domain templates: the Corr. Template includes the correlated-noise covariance, whereas the Non-Corr. Template assumes that the ACC noises are uncorrelated. By holding the signal response, instrumental auto spectra, segmentation, and analysis band fixed, we isolate the impact of the missing cross-spectral structure on the inferred parameters.

The results indicate that Non-Corr. Template  parameter shifts whenever correlated ACC noise is present in the data, whereas the Corr. Template generally remains closer to the injected values. The parameter absorbing the largest fraction of the mismatch depends on the spectral structure of the relevance with the correlated noise. For the WN prescription, the correlated component is most strongly degenerate with the SGWB spectral slope, while for the PS prescription, it is most strongly degenerate with the ACC-noise amplitude.
These results provide a quantitative characterization of how correlated ACC noise can bias parameter inference for specific correlation models and amplitudes. Such a characterization can help establish expectations for the magnitude and parameter dependence of this systematic in future mission data, potentially enabling correlated noise to be identified and distinguished from other contributions more efficiently and reliably.

The remainder of this paper is organized as follows. In Section~\ref{sec:theo_framework} we introduce the single-link data model, the models of the correlated ACC noise, and the mapping from single-link spectra to the TDI covariance. In Section~\ref{sec:PE} we present the $AE$-channel likelihood framework and discuss the parameter-estimation results obtained with correlated and non-correlated templates. We summarize the implications of these results in Section~\ref{sec:conclusion}.

\section{
Signal and Noise Modeling } \label{sec:theo_framework}

This section develops the forward model required to quantify the bias produced when correlated ACC noise is present in the data but absent from the analysis template. The construction proceeds in two stages. We first define the single-link observables and specify the auto- and cross-spectral structure of their instrumental noise. We then propagate the SGWB, ACC, OMS, and correlated ACC components through the moving-orbit TDI response to obtain the covariance used in parameter inference.

\subsection{Single-Link Data Model and Correlated Acceleration-Noise Prescription}

\begin{figure}[tbp]
    \centering
    \includegraphics[width=15cm, height=12cm]{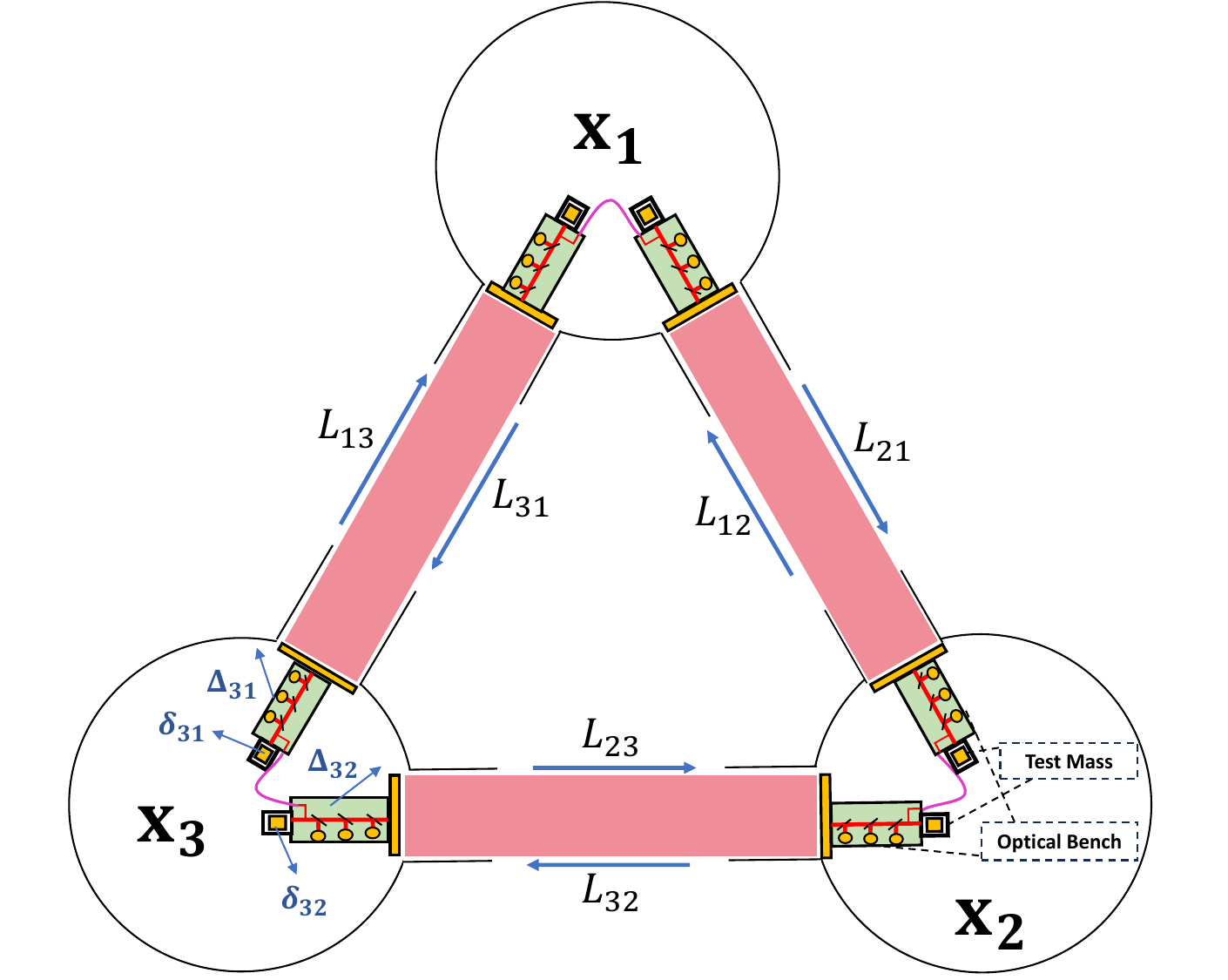}
    \caption{Taiji constellation and single-link notation, where the three spacecrafts are labeled as $\vb{x}_1$, $\vb{x}_2$ and $\vb{x}_3$. The arrows along the arms indicate the six directed optical paths $\boldsymbol{L}_{ij}$ from emitting spacecraft $\vb{x}_j$ to receiving spacecraft $\vb{x}_i$. The jitter motions of the TMs and OBs are denoted by $\boldsymbol{\delta}_{ij}$ and $\boldsymbol{\Delta}_{ij}$, with indices directed from the local to the distant spacecraft.}
    \label{fig:scheme}
\end{figure}

We begin with the standard single-link description of a space-based laser interferometer. As shown in Fig.~\ref{fig:scheme}, the constellation contains three spacecrafts at positions $\vb{x}_1$, $\vb{x}_2$, and $\vb{x}_3$, connected by six directed optical links. Gravitational waves (GWs) are measured through the phase fluctuations accumulated by laser light propagating along these links. Because the constellation orbits and rotates, the arm lengths are generally unequal and time dependent. Each spacecraft carries two movable optical subassemblies (MOSAs) that point toward the other spacecraft. A MOSA contains an optical bench (OB) and a nearby free-falling TM that provides the local inertial reference. In the ideal drag-free limit, the TM follows geodesic motion, while non-ideal effects, including thermal disturbances, residual-gas pressure, magnetic couplings, and control imperfections, give rise to a jitter noise $\boldsymbol{\delta}_{ij}$. The optical bench is similarly displaced by spacecraft motion and local disturbances, described by $\boldsymbol{\Delta}_{ij}$. The directed-index and jitter noise conventions defined in Fig.~\ref{fig:scheme} are adopted throughout the subsequent single-link and TDI calculations.

The GW-induced phase fluctuations are accompanied by additional contributions from laser, test-mass, and optical-bench motion. Each OB therefore provides three interferometric observables: the science interferometer $s_{ij}$, the test-mass interferometer $\tau_{ij}$, and the reference interferometer $r_{ij}$. These observables supply the information needed to cancel optical-bench motion and, after TDI, suppress the dominant laser-frequency noise. As an explicit example, the observables associated with $\text{MOSA}_{32}$ are \cite{2023PhRvD.108h2004Q,Otto:2015erp}
\begin{subequations}\label{eq:ifi}
    \begin{align}
        s_{32}&=y_{32}+D_{32}p_{23}-p_{32}-\frac{1}{c}\vb{\hat{n}_{23}}\cdot (D_{32}\vb{\Delta_{23}}-\vb{\Delta_{32}})+N_{32}^s, \\
        \tau_{32}&=p_{31}-p_{32}-\frac{2}{c}\vb{\hat{n}_{23}}\cdot (\boldsymbol{\delta_{32}}-\vb{\Delta_{32}})+N_{32}^\tau, \\
        r_{32}&=p_{31}-p_{32}+N_{32}^r.
    \end{align}
\end{subequations}
The other MOSA measurements follow by cyclic permutation of the indices. Throughout this work, the ordered pair $ij$ denotes a link received at spacecraft $i$ and emitted from spacecraft $j$. Accordingly, $\boldsymbol{\hat{n}}_{ij}$ is the unit vector between two spacecrafts, while $y_{ij}(t)$ and $p_{ij}$ denote the GW-induced Doppler response and laser-frequency noise, respectively. The arm vector shown in Fig.~\ref{fig:scheme} can be written as $\boldsymbol{L}_{ij}(t) = cL_{ij}(t)\vb{\hat n}_{ij}(t)$,
where $cL_{ij}(t)=|\boldsymbol{L}_{ij}(t)|$ is the physical arm length and ${L}_{ij}(t)$ is the associated light-travel time.
With this convention, the SGWB contribution to the single-link Doppler response is \cite{Baghi_2023}
\begin{equation}
	y_{ij}(t) =\sum_{\vb{\hat{k}}} \frac{1}{2\left[1 - \vb{\hat{k}} \cdot \vb{\hat{n}}_{ij}(t)\right]} \left[ H_{ij}\left(t - L_{ij}(t) - \mathbf{\hat{k}} \cdot \mathbf{x}_j(t)\right) - H_{ij}\left(t - \mathbf{\hat{k}} \cdot \mathbf{x}_i(t)\right) \right],
	\label{eq:single_link_response}
\end{equation}
where $H_{ij}\equiv\vb{\hat{n}}_{ij} \vb{h} \vb{\hat{n}}^\top_{ij}$ is the strain projected onto that arm and depends on the strain tensor $\vb{h}$ and the arm direction $\vb{\hat{n}}_{ij}$. $\vb{\hat{k}}$ denotes the direction of a GW source.
The corresponding time-delay operator is defined as
\begin{equation}
    D_{ij}(t)h(t)=h\left(t-L_{ij}(t)\right).
\end{equation}

The terms $N^s_{ij}$, $N^\tau_{ij}$, and $N^r_{ij}$ in Eqs.~\eqref{eq:ifi} represent the secondary interferometric noises, dominated by readout noise, optical-path noise, and fiber back-link noise. 
In unequal-arm space-based detectors, laser-frequency noise is many orders of magnitude larger than the target GW signal, and TDI is therefore employed to suppress this dominant noise contribution \cite{2021LRR....24....1T}. Before the delayed TDI combinations are formed, the elementary readouts are first combined so that the explicit optical-bench jitter noise terms cancel \cite{2023PhRvD.108h2004Q}.
After the pre-processing step, the measured Doppler data on a single laser link can be written as \cite{PhysRevD.107.123531}
\begin{equation}
	\eta_{ij}(t) = y_{ij}(t) + N^{\text{acc}}_{ij}(t) + D_{ij}(t) N^{\text{acc}}_{ji}(t) + N^{\text{oms}}_{ij}(t).
	\label{eq:total_data_link}
\end{equation}
Here we treat the laser-frequency noise $p_{ij}$ as perfectly removed by the subsequent TDI combinations. The remaining instrumental terms are ACC noise $N^{\text{acc}}_{ij}(t)$ and OMS noise $N^{\text{oms}}_{ij}(t)$. 
The former is the effective contribution associated with residual test-mass motion, while the latter combines the readout and optical-path noises that remain in the link observable. The ACC contribution in the single-link measurement is related to the test-mass jitter through
\begin{equation}
    N^{\text{acc}}_{ij}(t)=-\frac{1}{c}\vb{\hat{n}}_{ji}(t)\cdot \boldsymbol{\delta}_{ij}(t),
\end{equation}
which depends only on the motion of the local TM.

As discussed above, the residual acceleration of a nominally free-falling TM may originate from several environmental and instrumental sources. 
In particular, adjacent TMs are influenced by a common drag-free and attitude control system (DFACS) \cite{2017arXiv170200786A,2024arXiv240207571C}, which motivates allowing their ACC noise to be correlated. 
To capture this possibility, we consider a pair of Fourier-domain ACC-noise streams $\tilde{N}^{\text{acc}}_{ij}(f)$ and $\tilde{N}^{\text{acc}}_{ik}(f)$ in the same spacecraft, and write their covariance matrix as
\begin{equation}
        \expval{\left\{\tilde{N}^{\text{acc}}_{ij}(f),\tilde{N}^{\text{acc}}_{ik}(f)\right\}^\top\left\{\tilde{N}^{*\text{acc}}_{ij}(f),\tilde{N}^{*\text{acc}}_{ik}(f)\right\}}_{\text{ens}}=\frac{T_{\text{obs}}}{2} \begin{bmatrix}
S^{\text{acc}}_{ij}(f) & C^{\text{acc}}_{ij,ik}(f) \\
C^{\text{acc}}_{ik,ij}(f) & S^{\text{acc}}_{ik}(f) \\
\end{bmatrix}.
\label{eq:accm}
\end{equation}

To specify the diagonal entries of this covariance, we adopt the standard one-sided ACC-noise PSD \cite{PhysRevD.108.083007}
\begin{equation}
	S^{\text{acc}}_{ij}(f) = \left( \frac{s^{\text{acc}}_{ij}}{2\pi fc} \right)^2 \left[1 + \left(\frac{0.4~\text{mHz}}{f}\right)^2 \right] \left[1 + \left(\frac{f}{8~\text{mHz}}\right)^4 \right],
	\label{eq:acc_noise_PSD}
\end{equation}
with amplitude parameter $s^{\text{acc}}_{ij}$. Following the usual equal-noise reference configuration, we assume identical spectral shapes on all six links and adopt a common ACC-noise level $s^{\text{acc}}_{ij}=s_{\text{acc}} = 3 \times 10^{-15}~\text{m}/\text{s}^2/\sqrt{\text{Hz}}$. In this notation, $S^{\text{acc}}_{ij}(f)=S_{\text{acc}}(f)$ for every link.

Since the detailed frequency dependence of the correlated component is still uncertain, we consider two simple phenomenological models with distinct spectral shapes: a white-noise (WN) model and a PSD-scaled (PS) model. These two cases allow us to examine how the impact of correlated noise on parameter inference depends on its frequency dependence.
\begin{equation}
    C^{\text{acc}}_{ij,ik}(f) = C_{\text{acc}}^{\text{WN}}(f)=\rho\,\min_{f}\left[ S_{\text{acc}}(f)\right],
\label{eq:cacc_wn}
\end{equation}
\begin{equation}
    C^{\text{acc}}_{ij,ik}(f) = C_{\text{acc}}^{\text{PS}}(f)=\rho\,S_{\text{acc}}(f).
\label{eq:cacc_ps}
\end{equation}
The scaling parameter $\rho \in \{0.0,\, 0.1,...,\, 1.0\}$ controls the amplitude of $C_{\text{acc}}$. In the WN prescription, the normalization is chosen to ensure that the injected cross spectra do not exceed the corresponding auto spectra across the entire analysis band. In the PS prescription, the correlated component directly follows the frequency dependence of the ACC PSD. Measurements from LISA Pathfinder suggest that low-frequency ACC noise contains several physical contributions beyond a single idealized spectrum \cite{PhysRevLett.120.061101,PhysRevLett.116.231101}. These measurements do not directly establish the adjacent-test-mass correlations considered here. Nevertheless, we consider such correlations as an exploratory scenario to investigate their potential impact on SGWB and noise-parameter inference.

Having specified the link-level auto and cross spectra, we next propagate both the SGWB and the instrumental noise through the delayed single-link response before constructing the TDI covariance. Focusing first on the correlated ACC contribution, the relevant single-link spectra are
\begin{subequations}
    \label{eq:sig_link_acc}
    \begin{align}
        S^{\eta,\text{acc}}_{ij,ij}(f,t_m)&=2S_{\text{acc}}(f),\\
        S^{\eta,\text{acc}}_{ij,ji}(f,t_m)&=\left[\tilde{D}_{ij}(f,t_m)+\tilde{D}^*_{ji}(f,t_m)\right]S_{\text{acc}}(f), \\
        S^{\eta,\text{acc}}_{ij,ik}(f,t_m)&=C_{\text{acc}}(f), \\
        S^{\eta,\text{acc}}_{ij,ki}(f,t_m)&=\tilde{D}^{*}_{ki}(f,t_m)C_{\text{acc}}(f), \\
        S^{\eta,\text{acc}}_{ji,ik}(f,t_m)&=\tilde{D}_{ji}(f,t_m)C_{\text{acc}}(f), \\
        S^{\eta,\text{acc}}_{ji,ki}(f,t_m)&=\tilde{D}_{ji}(f,t_m)\tilde{D}^{*}_{ki}(f,t_m)C_{\text{acc}}(f),
    \end{align}
\end{subequations}
where $S^{\eta,\text{acc}}_{ij,lm}(f,t_m)$ denotes the cross-spectral density (CSD) between the ACC contributions entering two single-link observables,
\begin{equation}
    S^{\eta,\text{acc}}_{ij,lm}(f,t_m)=\frac{2}{T_c}\expval{\tilde{\eta}^{\text{acc}}_{ij}(f,t_m)\tilde{\eta}^{*\text{acc}}_{lm}(f,t_m)}_{\text{ens}}, \quad \tilde{\eta}^{\text{acc}}_{ij}(f,t_m)=\tilde{N}^{\text{acc}}_{ij}(f)+\tilde{D}_{ij}(f,t_m)\tilde{N}^{\text{acc}}_{ji}(f).
\end{equation}
Here the frequency-domain delay operator $\tilde{D}_{ij}(f,t_m) \equiv \mathcal{F}[D_{ij}(t,t_m)]$ is the Fourier transform of the corresponding time-domain operator. Under the moving-orbit configuration, the arm lengths vary in time and the delay operator is therefore non-stationary. 
We therefore divide the full time series, with total observation time $T_{\rm obs}$, into $N_c$ short segments of equal duration
\begin{equation}
T_c=\frac{T_{\rm obs}}{N_c},
\end{equation}
and treat the arm lengths as constant within each segment.  
The time point $t_m$ is chosen as the central time of the $m$-th bin, where $m \in [0,N_c-1]$. In each short bin, $\tilde{D}_{ij}(f,t_m)$ can then be computed as
\begin{equation}
    \tilde{D}_{ij}(f,t_m) \equiv e^{-i2\pi f L_{ij}(t_m)}.
\end{equation}
The OMS contribution is specified with the standard spectral shape \cite{PhysRevD.108.083007}
\begin{equation}
	S^{\text{oms}}_{ij}(f) = S_{\text{oms}}(f) = \left( s_{\text{oms}} \frac{2\pi f}{c} \right)^2 \left[1 + \left(\frac{2~\text{mHz}}{f}\right)^4 \right].
	\label{eq:oms_noise_PSD}
\end{equation}
We again use identical spectra on all six links and set $s_{\text{oms}}=8\times10^{-12}~\text{m}/\sqrt{\text{Hz}}$. As in the ACC sector, this equal-noise reference configuration removes the intrinsic OMS variation as a confounding effect.

We now introduce the SGWB signal component of the single-link measurement model. Assuming a stationary, Gaussian, isotropic, and unpolarized background, the SGWB is characterized by \cite{Allen_1999}
\begin{equation}
	\langle \tilde{h}_A(f;\mathbf{\hat{k}}) \tilde{h}^*_{A'}(f';\mathbf{\hat{k}'}) \rangle_{\text{ens}} = \frac{1}{16\pi} S_{\text{gw}}(f)\delta(f-f')\delta_{AA'}\delta_{\mathbf{\hat{k}}\mathbf{\hat{k}'}} ,
	\label{eq:GW_PSD_def}
\end{equation}
which implies
\begin{equation}
	\langle \tilde{h}(f) \tilde{h}^*(f) \rangle_{\text{ens}} = \frac{T_{\text{obs}}}{2} S_{\text{gw}}(f),
	\label{eq:GW_PSD_total}
\end{equation}
for the sky- and polarization-averaged strain. To connect this strain spectrum to the parameters inferred below, we use the dimensionless energy density
\begin{equation}
	S_{\text{gw}}(f) = \frac{3 q_0^2}{2\pi^2 f^3}\left[q^2 \Omega_{\text{gw}}(f)\right],
	\label{eq:GW_Omega_relation}
\end{equation}
where $q$ is the dimensionless Hubble parameter and $q_0=100~\text{km}\,\text{s}^{-1}\,\text{Mpc}^{-1}$. We model the SGWB energy-density spectrum as a power law
\begin{equation}
	q^2 \Omega_{\text{gw}}(f) = \Omega_0 \left( \frac{f}{f_0} \right)^\gamma,
	\label{eq:GW_power_law}
\end{equation}
with pivot $f_0=1~\text{mHz}$. All subsequent simulations use $\Omega_0=3\times10^{-12}$ and $\gamma=-1$.

Having specified the ACC-noise contribution in Eqs.~\eqref{eq:sig_link_acc}, we now give the OMS-noise and SGWB contributions to the single-link spectral matrix
\begin{equation}
    S^{\eta,\text{oms}}_{ij,lm}(f,t_m)=\left\{
\begin{aligned}
S_{\text{oms}}(f),\quad &\text{if}\ ij=lm \\
0,\quad &\text{else}
\end{aligned}
\right.
\label{eq:sig_link_oms}
\end{equation}

\begin{equation}
    S^{\eta,\text{gw}}_{ij,lm}(f,t_m) = \tilde{R}^{\eta,\text{gw}}_{ij,lm}(f,t_m) S_{\text{gw}}(f),
    \label{eq:sig_link_gw}
\end{equation}
where we consider no correlation among the OMS noise in different MOSAs. By contrast, the SGWB generally correlates different links through the response function $\tilde{R}^{\eta,\text{gw}}_{ij,lm}(f,t_m)$, which maps the strain spectrum into frequency-domain Doppler measurements. Combining the ACC, OMS, and SGWB terms gives the total single-link spectral matrix
\begin{equation}
    S^{\eta}_{ij,lm}(f,t_m) = \sum_{\alpha} S^{\eta, \alpha}_{ij,lm}(f,t_m) = \sum_{\alpha} \tilde{R}^{\eta,\alpha}_{ij,lm}(f,t_m) S_{\alpha}(f),
\end{equation}
where $\alpha \in \{\text{acc},\, \text{oms},\, \text{gw}\}$.

\subsection{TDI Observables and Frequency-Domain Spectra} \label{sec:TDIobservable}

The link-level spectra must next be mapped to TDI observables in which laser-frequency noise is suppressed. We begin with the unequal-arm Michelson combinations $X$, $Y$, and $Z$ \cite{Prince_2002,Tinto_2003}
\begin{equation}
	\begin{aligned}
		X(t) = & \left[\eta_{12}(t) + D_{12}\eta_{21}(t) - D_{13}D_{31}\eta_{12}(t) - D_{13}D_{31}D_{12}\eta_{21}(t)\right] \\
		       & - \left[\eta_{13}(t) + D_{13}\eta_{31}(t) - D_{12}D_{21}\eta_{13}(t) - D_{12}D_{21}D_{13}\eta_{31}(t)\right],
	\end{aligned}
	\label{eq:TDI_X}
\end{equation}
with $Y$ and $Z$ obtained by cyclic permutations of spacecraft indices. These variables are then transformed into the approximately orthogonal $A$, $E$, and $T$
\begin{equation}
	\begin{aligned}
		A(t) &= \frac{Z(t) - X(t)}{\sqrt{2}}, \\
		E(t) &= \frac{X(t) - 2Y(t) + Z(t)}{\sqrt{6}}, \\
		T(t) &= \frac{X(t) + Y(t) + Z(t)}{\sqrt{3}}.
	\end{aligned}
	\label{eq:TDI_AET}
\end{equation}
For an equal-arm constellation, the $AET$ channels are orthogonal, such that their cross spectra vanish. For the moving, unequal-arm constellation, however, the cross-spectral terms cannot in general be neglected. As quantified in Appendix~\ref{app:tdi_cross_power}, the normalized cross power $\abs{S_{UV}}/\sqrt{S_U S_V}$ shows that the $AT$ and $ET$ cross spectra are not uniformly negligible over the analysis band, whereas the $AE$ cross spectrum remains small. Including the full $AET$ covariance matrix in the likelihood would require evaluating additional cross-spectral terms at each likelihood evaluation, which is computationally expensive for nested sampling. We therefore restrict the parameter-estimation analysis to the $A$ and $E$ auto-spectra.

Applying Eq.~\eqref{eq:TDI_X} and its cyclic permutations, together with the $AET$ transformation in Eq.~\eqref{eq:TDI_AET}, and taking their Fourier transforms, each TDI channel can be expressed as a linear combination of the six frequency-domain single-link data streams
\begin{equation}
    \tilde{U}(f,t_m)=\sum_{ij}[\vb{w}_{U}(f,t_m)]_{ij}\tilde{\eta}_{ij}(f,t_m).
    \label{eq:TDI_to_siglink}
\end{equation}
Here, the $1\times6$ vector $\vb{w}_{U}(f,t_m)$ contains the transformation coefficients for $U\in\{A,E\}$.
The TDI cross spectrum then follows
\begin{equation}
    \begin{split}
        S_{UV}(f,t_m) &= \frac{2}{T_c}\expval{\tilde{U}(f,t_m)\tilde{V}^{*}(f,t_m)}_{\text{ens}}
        =\sum_{ij,lm}\left[\mathbf{W}_{UV}(f,t_m)\right]_{ij,lm}S^{\eta}_{ij,lm}(f,t_m) \\
        &=\sum_{ij,lm}\left[\mathbf{W}_{UV}(f,t_m)\right]_{ij,lm} \sum_{\alpha}\tilde{R}^{\eta,\alpha}_{ij,lm}(f,t_m) S_{\alpha}(f),
    \end{split}
    \label{eq:TDI_CSD_from_link_a}
\end{equation}
where we define $\left[\mathbf{W}_{UV}(f,t_m)\right]_{ij,lm}=[\vb{w}_{U}(f,t_m)]_{ij}[\vb{w}^*_{V}(f,t_m)]_{lm}$. Because the propagation is linear, the contribution from each physical component can be retained separately. We therefore define a source-specific transfer function from an input spectrum $S_\alpha(f)$ to the TDI output
\begin{equation}
    \tilde{R}^{\alpha}_{UV}(f,t_m)=\sum_{ij,lm}\left[\mathbf{W}_{UV}(f,t_m)\right]_{ij,lm}\tilde{R}^{\eta,\alpha}_{ij,lm}(f,t_m).
\end{equation}

The instrumental noise and SGWB contributions can then be propagated separately
\begin{equation}
    S^{\text{n}}_{UV}(f,t_m)=\tilde{R}^{n}_{UV}(f,t_m)S_{n}(f),
    \qquad
    S^{\text{gw}}_{UV}(f,t_m)=\tilde{R}^{\text{gw}}_{UV}(f,t_m)S_{\text{gw}}(f),
    \label{eq:TDI_CSD_signal_noise}
\end{equation}
where $n$ is an abbreviated label for the combined ACC and OMS contributions.

For intuition and as an analytic consistency check, we also quote the noise PSDs in the reference equal-arm limit
\begin{align}
        S^{n}_{A}(f)=S^{n}_{E}(f)=&16\sin^2{(\omega L)}\left\{ 3S_{\text{acc}}(f) + \left[2\cos{(\omega L)} + \cos{(2\omega L)}\right]\left[S_{\text{acc}}(f)-C_{\text{acc}}(f)\right] \right\} \nonumber \\
        &+8\sin^2{(\omega L)}\left[2 + \cos{(\omega L)}\right]S_{\text{oms}}(f),
    \label{eq:equal_arm_noise_psd}
\end{align}
where we define $\omega \equiv 2\pi f$. In this limit, the cross-spectral densities between the $A$ and $E$ channels vanish. 

We generate time-domain realizations of the injected spectra with the Taiji Data Challenge II toolkit \textbf{Triangle-Simulator} \cite{Du:2025xdq}. The one-year data set is nonstationary because the orbital response evolves continuously. We therefore divide it into 146 non-overlapping segments of 2.5 days, within each of which the response is treated as quasi-stationary. A Kaiser window with $\beta=28$ suppresses spectral leakage, and a short-time Fourier transform (STFT) is employed to perform the spectral analysis for each segment.

\begin{figure}[tbp]
    \centering
	\includegraphics[width=\linewidth]{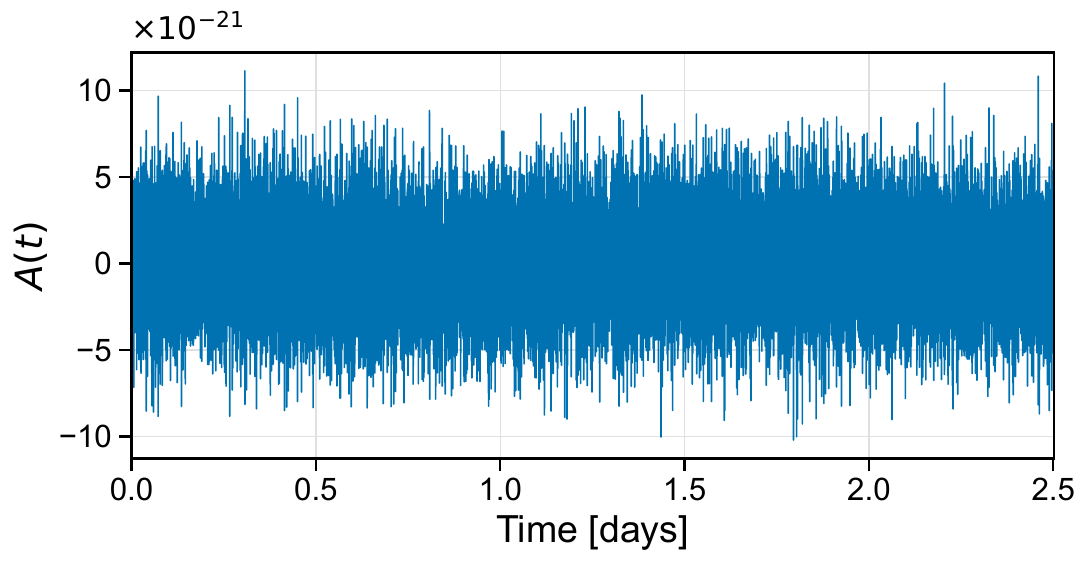}
	\captionof{figure}{Representative time series of channel $A$ for the WN model with $\rho=0.5$ over one 2.5-day segment. The blue solid curve denotes $A(t)$.}
	\label{fig:wn_rho05_time_series}
\end{figure}

\begin{figure}[tbp]
    \centering
	\includegraphics[width=\linewidth]{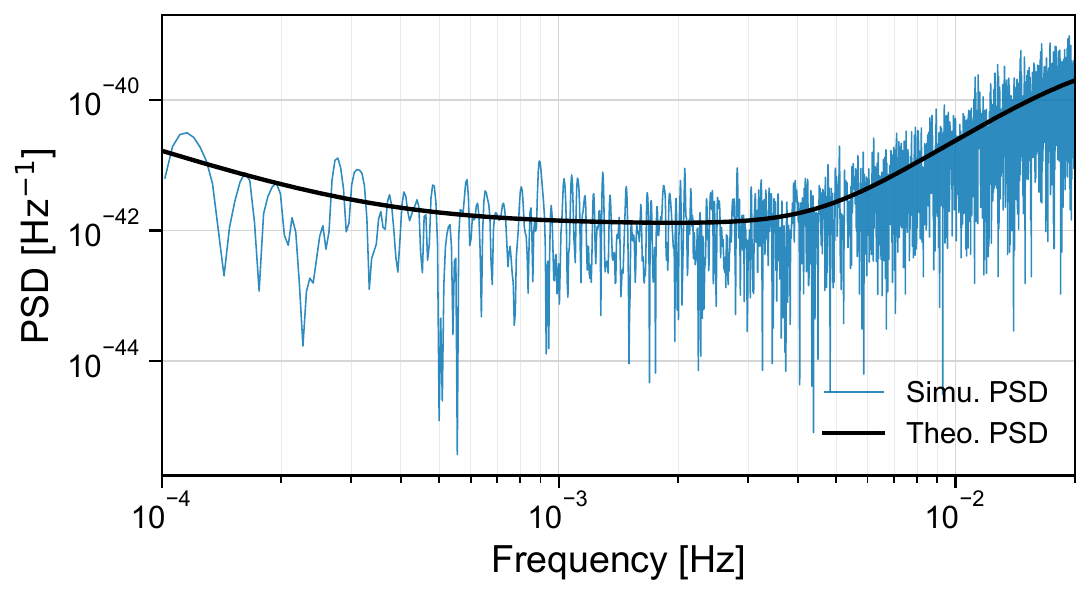}
	\captionof{figure}{PSD for the WN model with $\rho=0.5$ for channel $A$. The blue curve gives the PSD of the same 2.5-day segment shown in Fig.~\ref{fig:wn_rho05_time_series}; the black solid curve gives the theoretical total PSD for that segment. The segment PSD is estimated with a Kaiser window with $\beta=28$, and both curves are shown over the analysis band $\left[10^{-4},2\times10^{-2}\right]\,\mathrm{Hz}$.}
	\label{fig:wn_rho05_psd}
\end{figure}

We use the WN injection with $\rho=0.5$ as a representative diagnostic of the simulated data and take the $A$ channel as an example. The 2.5-day realization in Fig.~\ref{fig:wn_rho05_time_series} remains visually noise-like. Neither the SGWB nor the correlated ACC component produces a deterministic feature that can be separated in the time series. This is expected for two stochastic contributions and motivates characterizing their effects through the spectral covariance. Fig.~\ref{fig:wn_rho05_psd} shows the PSD of the same segment together with the theoretical total spectrum evaluated for the corresponding orbital configuration. We evaluate the total PSD over $[10^{-4}, 2 \times 10^{-2}]~\text{Hz}$, the band where the response is stable and the detector is most sensitive \cite{babak2021lisasensitivitysnrcalculations}. The figure shows that the simulated PSD follows the theoretical spectrum overall, while exhibiting fluctuations around the theoretical curve, as expected for a finite-duration stochastic realization. The overall agreement confirms that the injected correlated noise is consistent with the prescribed spectral model.

To quantify the relative strength of the SGWB signal against the correlated-noise background across the injected values of $\rho$, we use the network signal-to-noise ratio (SNR) in the $AE$ subspace \cite{Caprini_2019}
\begin{equation}
	\text{SNR} = \sqrt{T_c \sum_{t_m}\int_{f_{\text{min}}}^{f_{\text{max}}}\left[\left(\frac{S^{\text{gw}}_{A}(f,t_m)}{S^{n}_{A}(f,t_m)}\right)^2+\left(\frac{S^{\text{gw}}_{E}(f,t_m)}{S^{n}_{E}(f,t_m)}\right)^2\right]\dd{f}}.
	\label{eq:SNR_definition}
\end{equation}

\section{Parameter Inference} \label{sec:PE}

We compare two frequency-domain models to isolate the effect of omitting correlated ACC noise when it is present in the data. The Corr. Template retains $C_{\rm acc}(f)$ in the TDI covariance and therefore predicts the same covariance structure used to generate the mock data. In contrast, the Non-Corr. Template sets this contribution to zero while keeping the same SGWB response and instrumental auto-noise model. Any systematic differences between the resulting posteriors can thus be attributed specifically to the omitted covariance term, with the data, signal response, and remaining noise prescription held fixed.

\subsection{Likelihood function} \label{sec:likelihood_function}
We perform inference with the data vector $\tilde{\mathbf{d}}_{AE}(f,t_m)=\left[\tilde{A}(f,t_m), \ \tilde{E}(f,t_m)\right]^\top$. For each model, we infer the corresponding model parameters using Bayesian analysis. The posterior distribution is given by
\begin{equation}
	p\left(\boldsymbol{\theta}|\tilde{\mathbf{d}}_{AE}\right)\propto \mathcal{L}\left(\tilde{\mathbf{d}}_{AE}|\boldsymbol{\theta}\right)p(\boldsymbol{\theta}),
\end{equation}
where $\mathcal{L}\left(\tilde{\mathbf{d}}_{AE}|\boldsymbol{\theta}\right)$ is the likelihood and $p(\boldsymbol{\theta})$ is the prior. The two models differ in dimensionality because the Corr. Template must infer the correlation amplitude. Its sampled parameter vector is
\begin{equation}
    \boldsymbol{\theta}_{\rm cor}=\left\{\log_{10}\Omega_0,\gamma,s_{\rm oms},s_{\rm acc},\rho\right\}.
\end{equation}
Since the Non-Corr. Template fixes this contribution to zero, its sampled vector reduces to
\begin{equation}
    \boldsymbol{\theta}_{\rm nocor}=\left\{\log_{10}\Omega_0,\gamma,s_{\rm oms},s_{\rm acc}\right\}.
\end{equation}
The parameters common to both templates are assigned identical prior distributions, while $\rho$ is sampled only in the Corr. Template. The adopted prior ranges are summarized in Table~\ref{tab:prior}.
\begin{table}
\begin{center}
\caption{Prior distributions adopted in the parameter estimation. Here, $\log_{10}\Omega_0$ and $\gamma$ characterize the amplitude and spectral slope of the power-law SGWB in  Eq.~\eqref{eq:GW_power_law}, while $s_{\text{oms}}$ and $s_{\text{acc}}$ set the amplitudes of the OMS and ACC noise spectra in Eqs.~\eqref{eq:oms_noise_PSD} and~\eqref{eq:acc_noise_PSD}, respectively. The parameter $\rho$, included only in the Corr. Template, controls the strength of the correlated acceleration-noise component defined in Eqs.~\eqref{eq:cacc_wn} and \eqref{eq:cacc_ps} for the WN and PS prescriptions. Here $\mathcal{U}\left[\text{min}, \text{max}\right]$ denotes a uniform prior over the interval $\left[\text{min}, \text{max}\right]$.}
\begin{tabular}{lccr}
  \hline\noalign{\smallskip}
$\quad$&Parameters & Prior&$\quad$ \\
\hline
$\quad$&$\log_{10}\Omega_0$ & $\mathcal{U}\left[-15, -5 \right]$&$\quad$ \\
$\quad$&$\gamma$ & $\mathcal{U}\left[-3, 3 \right]$&$\quad$ \\
$\quad$&$s_{\text{oms}}~\text{pm}/\sqrt{\text{Hz}}$ & $\mathcal{U}\left[5, 20\right]$&$\quad$ \\
$\quad$&$s_{\text{acc}}~{\text{fm}}/{\text{s}^2}/\sqrt{\text{Hz}}$ & $\mathcal{U}\left[0, 10\right]$&$\quad$ \\
$\quad$&$\rho$ & $\mathcal{U}\left[0, 1\right]$&$\quad$ \\
\hline\noalign{\smallskip}
\end{tabular}
\label{tab:prior}
\end{center}
\end{table} 
Considering that the TDI data in each frequency bin are well described by a zero-mean complex Gaussian distribution, the likelihood can be written as \cite{Karnesis_2020}
\begin{equation}
    \begin{split}
        \ln\mathcal{L}\left(\tilde{\mathbf{d}}_{AE}|\boldsymbol{\theta}\right)&=-\sum_{f_k,t_m}\left[\tilde{\mathbf{d}}_{AE}^{\dagger}(f_k,t_m)\,\mathbf{M}^{-1}(f_k,t_m;\boldsymbol{\theta})\,\tilde{\mathbf{d}}_{AE}(f_k,t_m)+\ln\abs{\mathbf{M}(f_k,t_m;\boldsymbol{\theta})}\right] \\
        &\simeq-\sum_{f_k,t_m} \sum_{U}\left[\frac{\hat{S}_{U}(f_k,t_m)}{S_{U}(f_k,t_m;\boldsymbol{\theta})}+\ln\abs{S_{U}(f_k,t_m;\boldsymbol{\theta})}\right],
    \end{split}
\end{equation}
where $f_k$ denotes the discrete frequency points, and $\mathbf{M}$ is the model covariance of the selected TDI data. The first line displays the multivariate complex-Gaussian form. The second line is its diagonal Whittle approximation \cite{Romano:2016dpx,Adams_2010,Wang_2022}, in which $\hat{S}_U(f_k,t_m)$ is the measured auto spectrum. With the normalization in Eq.~\eqref{eq:GW_PSD_total}, the $AE$ covariance is
\begin{equation}
	\mathbf{M}(f_k,t_m)=\frac{T_c}{2}
	\begin{bmatrix}
		S_{A} & S_{AE} \\
		S_{EA} & S_{E}
	\end{bmatrix}
    \simeq \frac{T_c}{2}
    \begin{bmatrix}
		S_{A} & 0 \\
		0 & S_{E}
	\end{bmatrix}.
	\label{eq:covm_ae}
\end{equation}
In this formulation, all information relevant to the SGWB and instrumental noise is encoded in the frequency-dependent covariance matrix. It is therefore important to ensure that the contributions to the mock data and to the theoretical covariance entering the likelihood are treated consistently. Since the present study focuses on correlated ACC noise, we demonstrate this consistency explicitly for the $C_{\rm acc}(f)$ contribution in Appendix~\ref{app:cacc_analysis}. In particular, the relevant terms in the mock-data TDI auto spectra depend on $\text{Re}[\hat{C}^{\text{acc}}_{ij,ik}(f)]$, whose ensemble expectation is the prescribed cross spectrum $C_{\rm acc}(f)$. Thus, the correlated-noise contribution in the mock data and that in the theoretical covariance used in the likelihood agree in expectation. The same construction applies to the other signal and noise components.

Posterior sampling is then performed with nested sampling, which efficiently explores the correlated parameter space and provides a stable characterization of the posterior distributions for both templates.

\subsection{Inference Results}
\begin{figure}[t]
	\centering
	\includegraphics[width=0.495\textwidth]{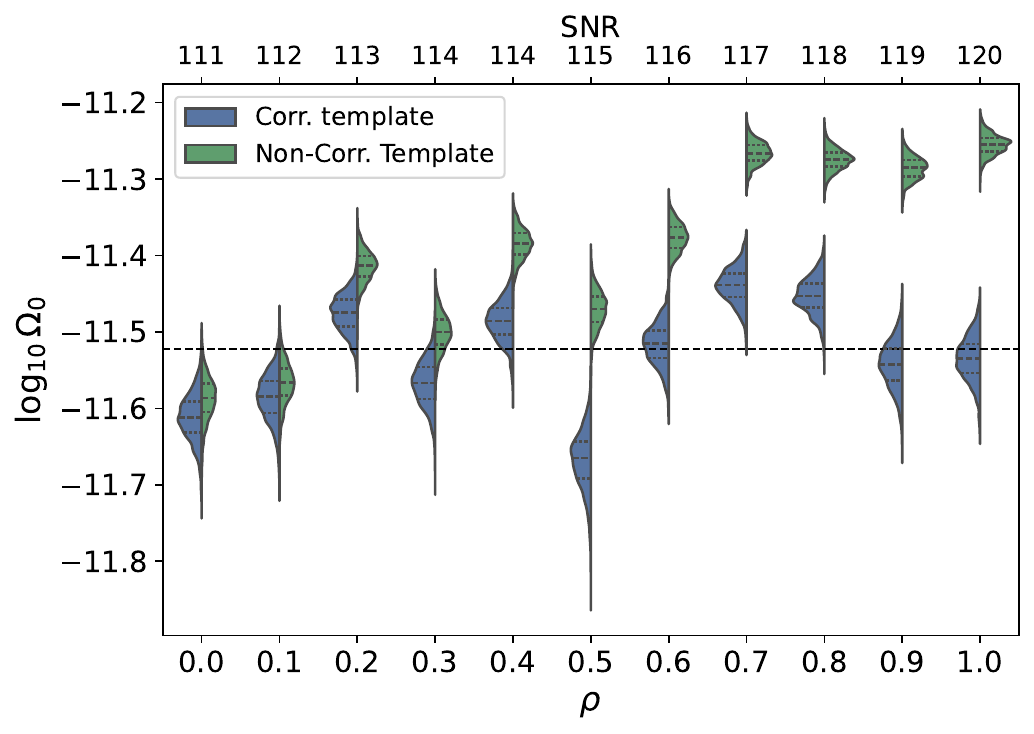}
	\hfill
	\includegraphics[width=0.495\textwidth]{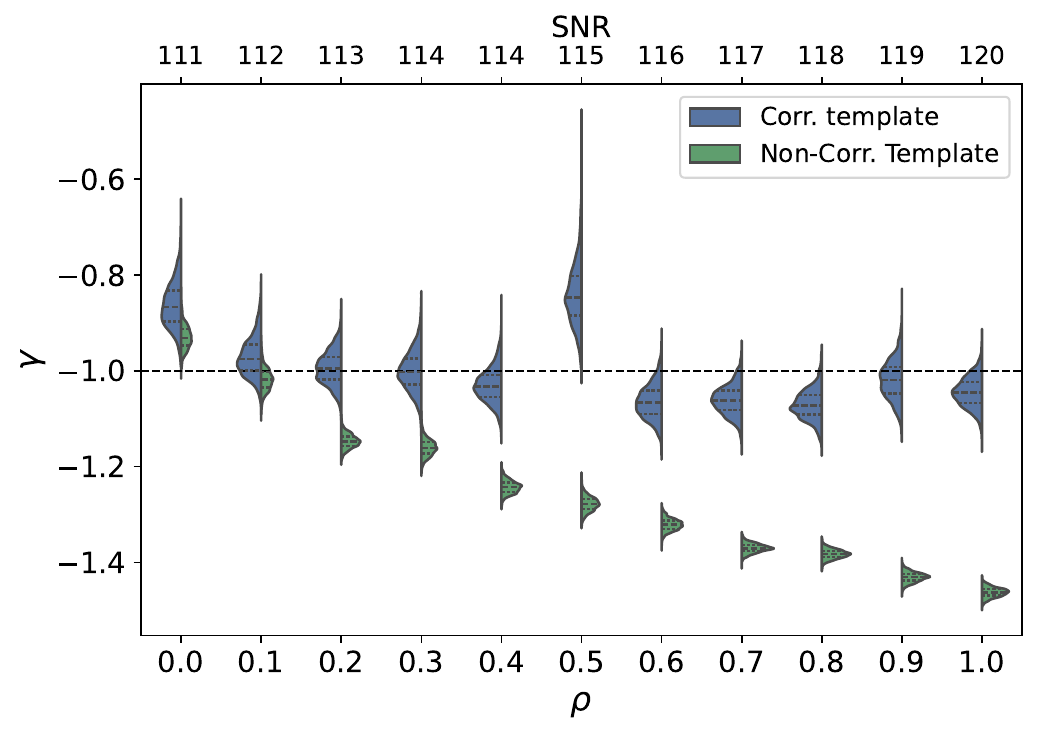}

	\vspace{0.3em}

	\includegraphics[width=0.495\textwidth]{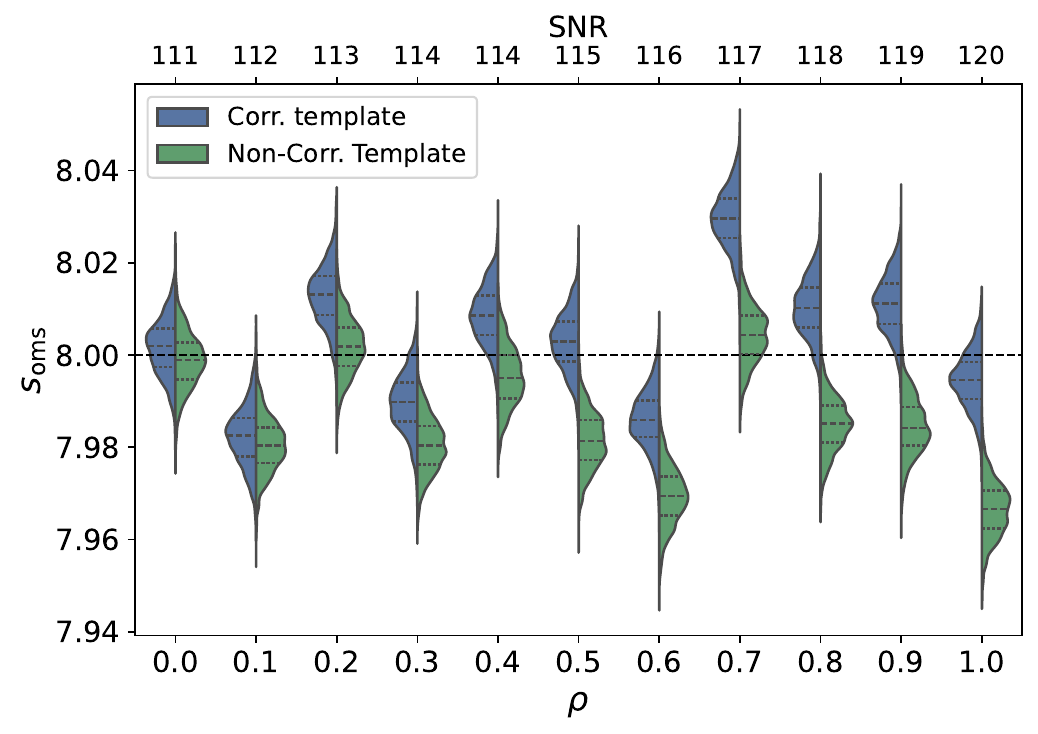}
	\hfill
	\includegraphics[width=0.495\textwidth]{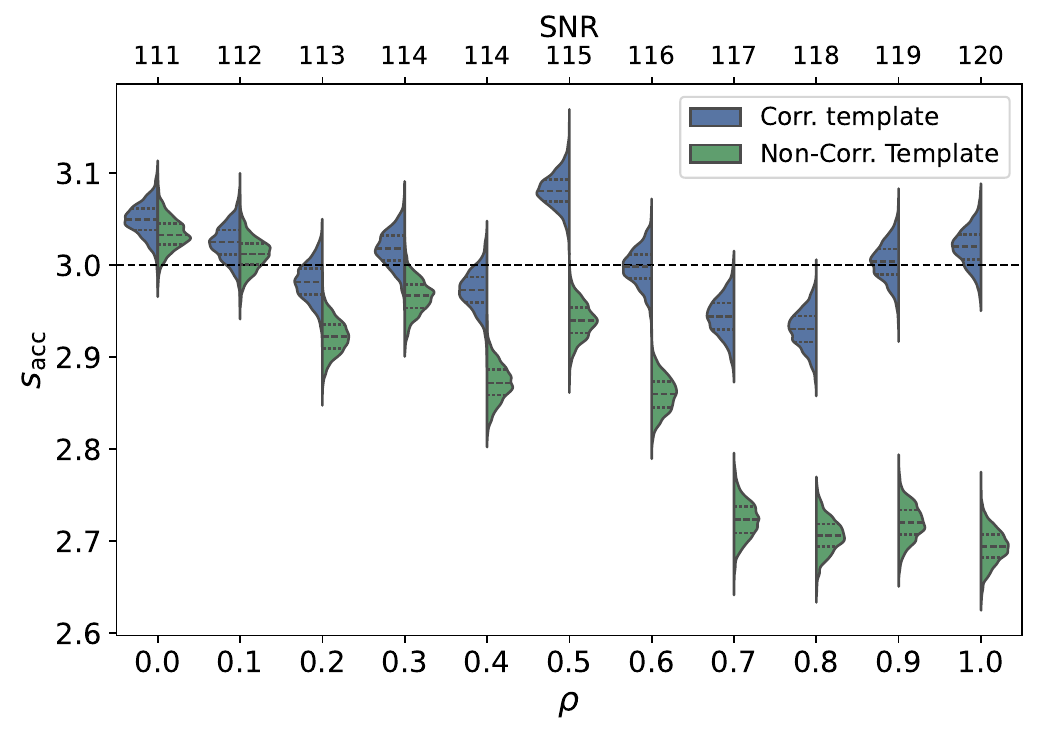}
	\caption{Marginalized $AE$-channel posteriors for the WN model in Eq.~\eqref{eq:cacc_wn}. The panels show $\log_{10}\Omega_0$ (top left), $\gamma$ (top right), $s_{\rm oms}$ (bottom left), and $s_{\rm acc}$ (bottom right). Blue and green half-violins denote the Corr. and Non-Corr. Templates, respectively. The internal black bars mark posterior quartiles, and the horizontal black dashed line marks the injected value. The lower axis gives the injected correlation $\rho$, and the upper axis gives the corresponding network SNR.}
	\label{fig:violin_wn}
\end{figure}

We assess the impact of the omitted covariance through the posterior distributions. For the WN prescription, Fig.~\ref{fig:violin_wn} shows how the inferred SGWB and instrumental-noise parameters respond when the covariance model is complete or incomplete. The Corr. Template additionally samples $\rho$, but this fifth coordinate is not shown in the violin overview because it has no counterpart in the Non-Corr. Template. Across most of the scanned values of $\rho$, the posterior median of the Corr. Template remains closer to the injections in all four shared parameters. This indicates that once the correlated contribution is explicitly retained in the Corr. model, the fit no longer needs to absorb that extra power by shifting the SGWB and instrumental-noise parameters away from their injected values. In contrast, the Non-Corr. Template has no such degree of freedom and must redistribute the missing contribution among other parameters. As $\rho$ increases, the omitted contribution becomes larger, and the resulting biases in the shared parameters generally become more pronounced. 

\begin{figure}[t]
	\centering
	\includegraphics[width=0.495\textwidth]{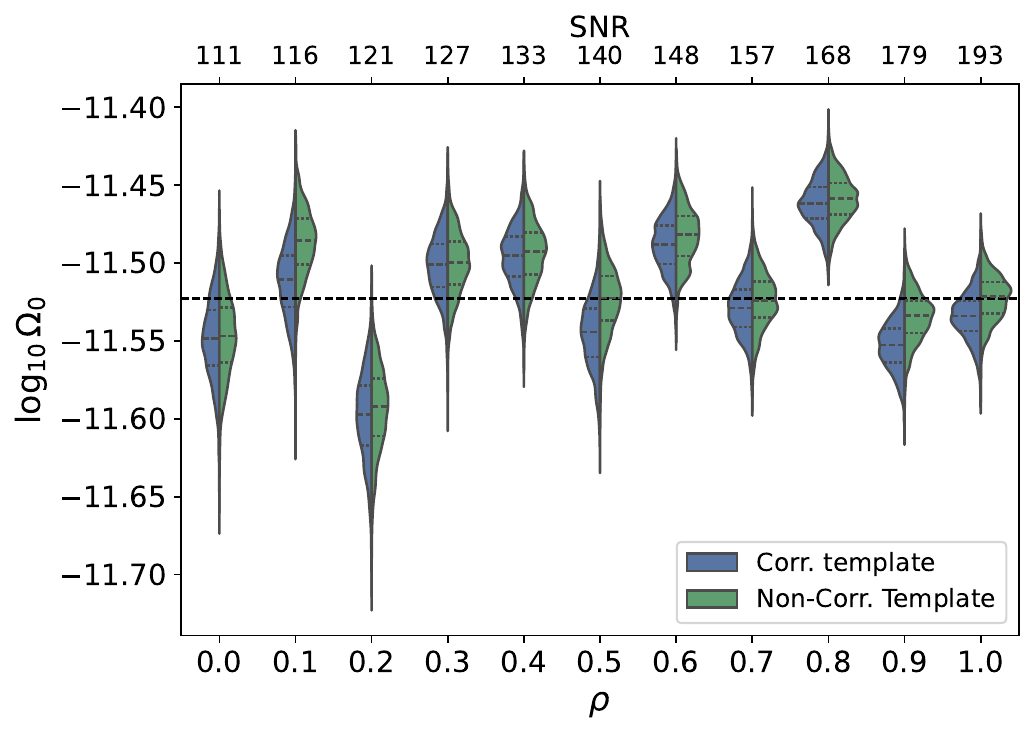}
	\hfill
	\includegraphics[width=0.495\textwidth]{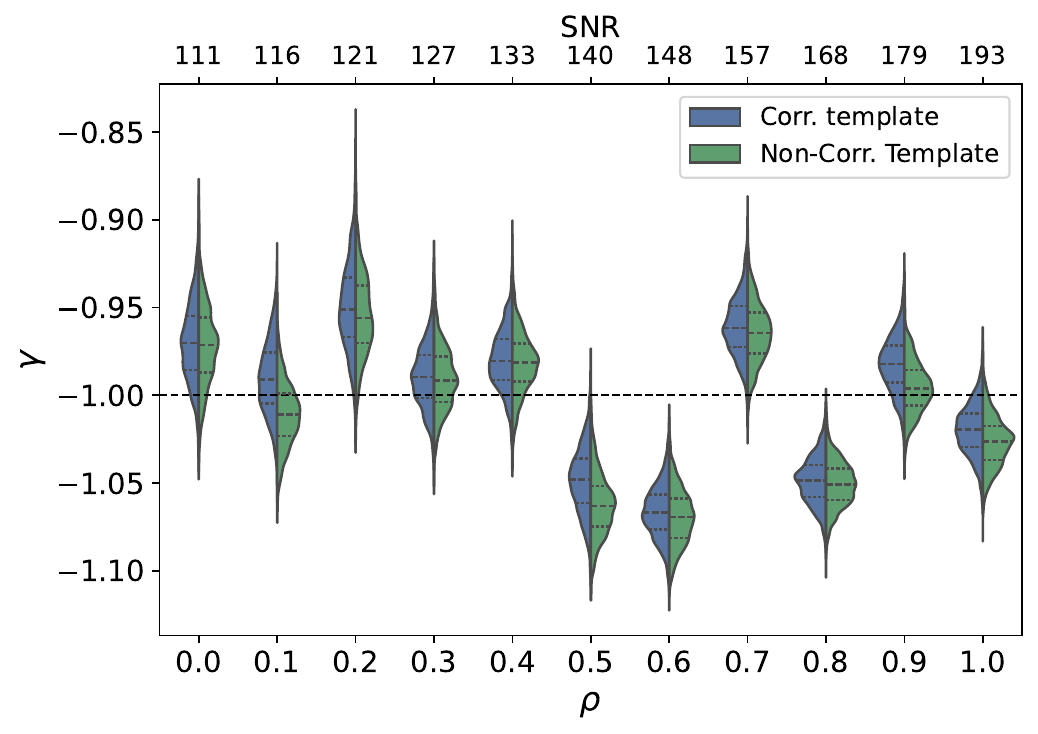}

	\vspace{0.3em}

	\includegraphics[width=0.495\textwidth]{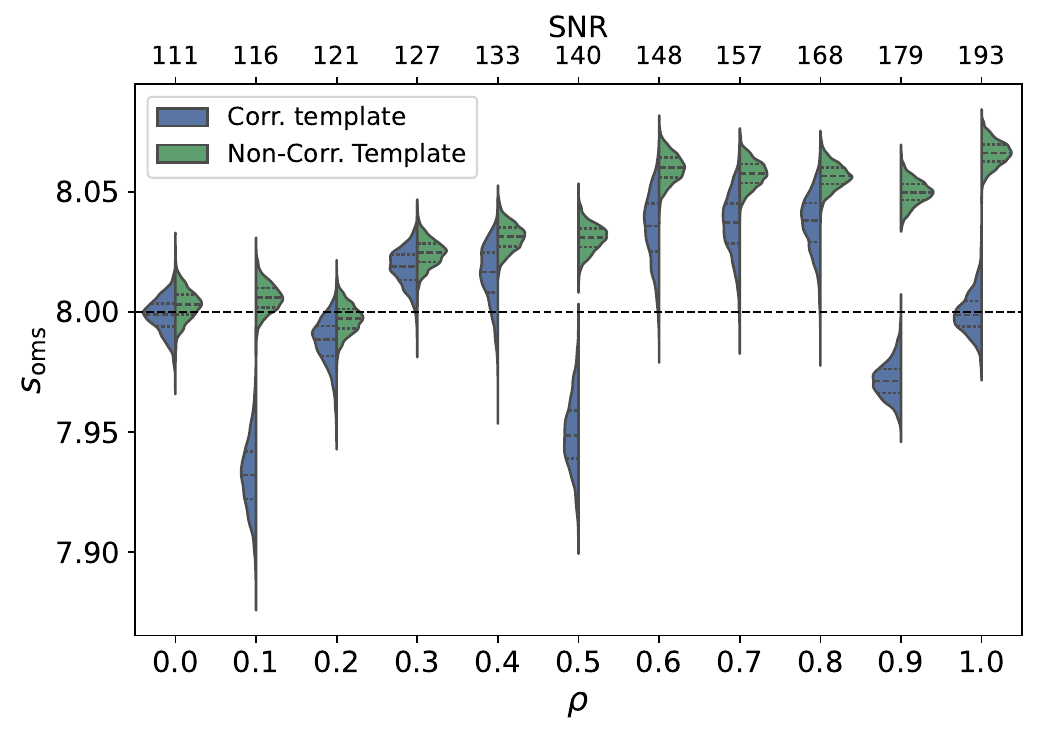}
	\hfill
	\includegraphics[width=0.495\textwidth]{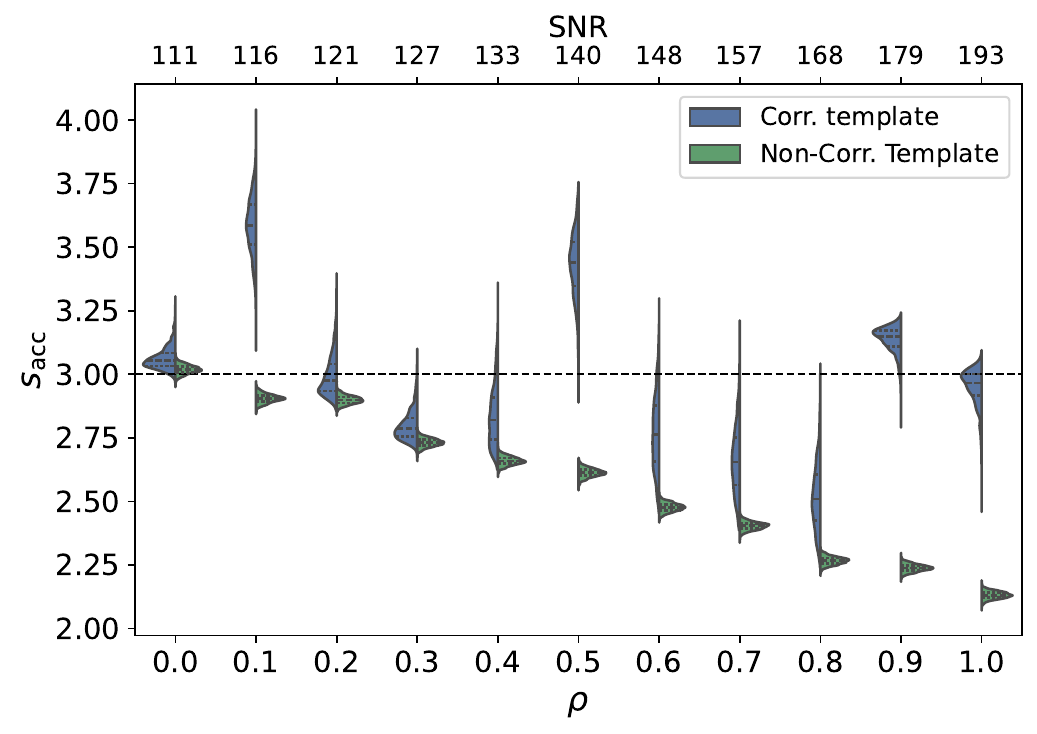}
	\caption{Marginalized $AE$-channel posteriors for the PS model in Eq.~\eqref{eq:cacc_ps}. The panels show $\log_{10}\Omega_0$ (top left), $\gamma$ (top right), $s_{\rm oms}$ (bottom left), and $s_{\rm acc}$ (bottom right). Blue and green half-violins denote the Corr. and Non-Corr. Templates, respectively. The internal black bars mark posterior quartiles, and the horizontal black dashed line marks the injected value. The lower axis gives the injected correlation $\rho$, and the upper axis gives the corresponding network SNR.}
	\label{fig:violin_ps}
\end{figure}

We next turn to the PS model, shown in Fig.~\ref{fig:violin_ps}. The qualitative outcome of the template comparison is unchanged: the Non-Corr. Template develops systematic offsets when the injected correlation becomes appreciable, whereas the Corr. Template remains closer to the injected values overall.

As seen in Figs.~\ref{fig:violin_wn} and~\ref{fig:violin_ps}, the largest systematic shift occurs in different parameters for the two noise models: $\gamma$ for the WN model and $s_{\rm acc}$ for the PS model, reflecting the different spectral shapes of $C_{\rm acc}^{\rm WN}(f)$ and $C_{\rm acc}^{\rm PS}(f)$.
To quantify this behavior, we evaluate the expected Fisher matrix for the same diagonal $AE$ Whittle likelihood used in the inference pipeline. For each injected value of $\rho$, the Fisher matrix is evaluated at the corresponding fiducial parameter point. It is given by
\begin{equation}
    F_{\mu\nu}
    =\sum_{t_m, f_k}\sum_{U}
    \frac{\partial_\mu S_U(f_k,t_m;\boldsymbol{\theta})\,\partial_\nu S_U(f_k,t_m;\boldsymbol{\theta})}
    {S_U^2(f_k,t_m;\boldsymbol{\theta})},
\end{equation}
where $\mu$ and $\nu$ index the parameters. Within the usual local Gaussian approximation, the corresponding parameter-correlation matrix coefficients (PCCs) are
\begin{equation}
    \mathrm{PCC}_{\mu\nu}=\frac{F^{-1}_{\mu\nu}}{\sqrt{F^{-1}_{\mu\mu}F^{-1}_{\nu\nu}}}.
\end{equation}

\begin{figure}[t]
	\centering
	\includegraphics[width=0.48\textwidth]{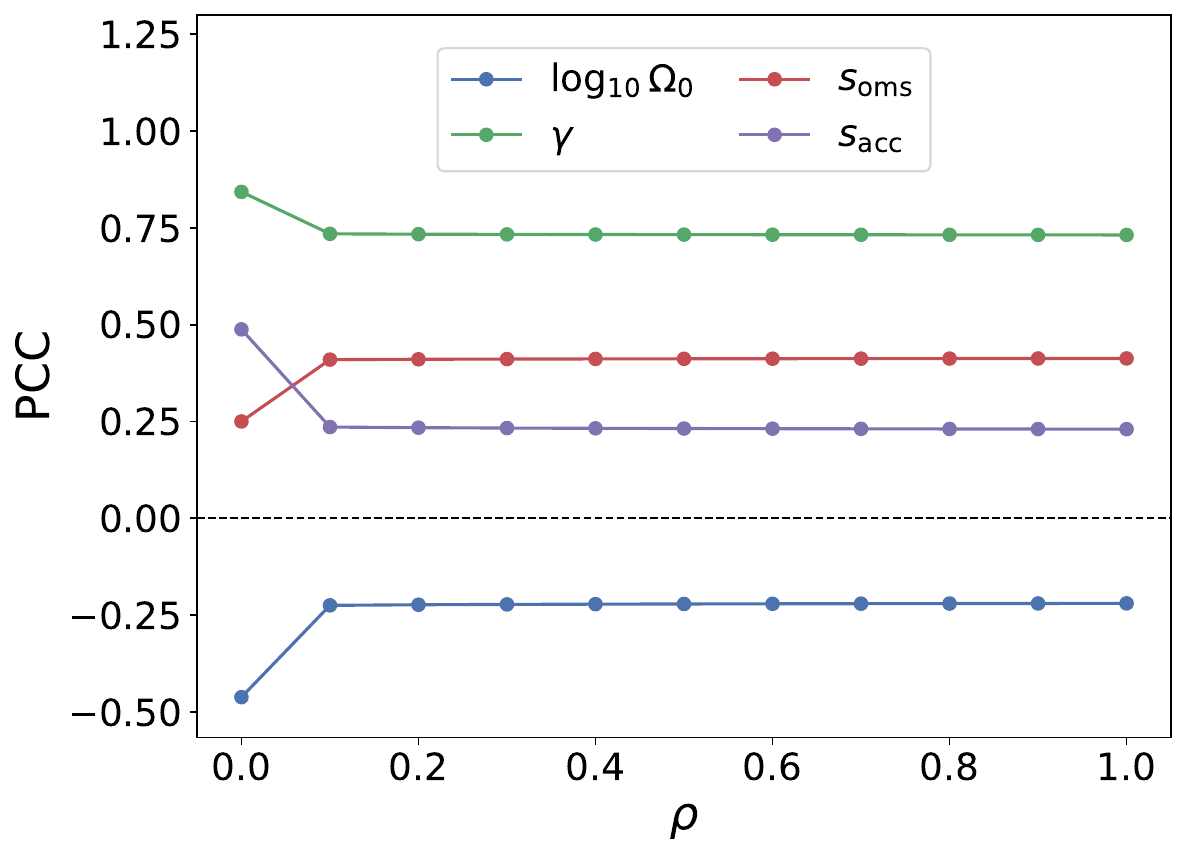}
	\includegraphics[width=0.48\textwidth]{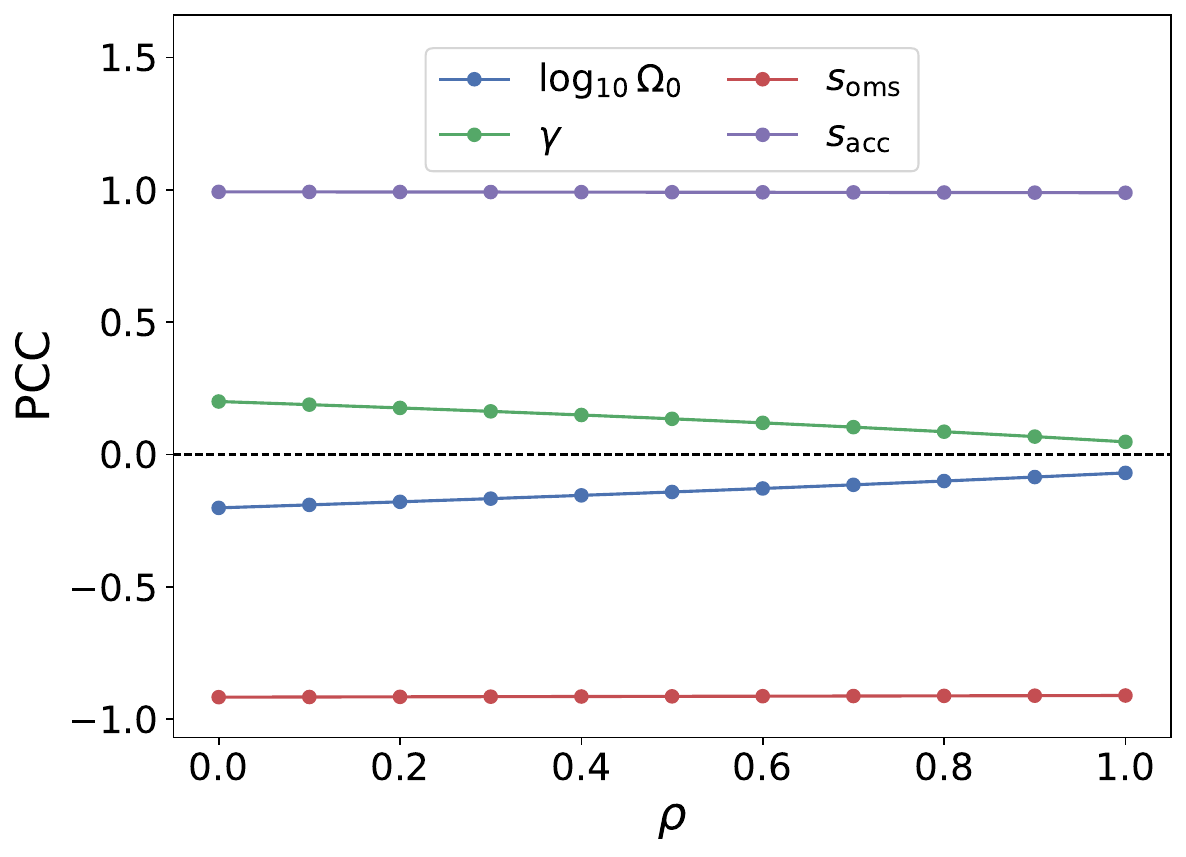}
	\caption{Fisher PCCs between $\rho$ and the remaining parameters for the WN model (left) and PS model (right), evaluated with the diagonal $AE$ likelihood. Blue, green, red, and purple solid lines denote $\log_{10}\Omega_0$, $\gamma$, $s_{\rm oms}$, and $s_{\rm acc}$, respectively. The horizontal black dashed line marks zero correlation.}
	\label{fig:fisher}
\end{figure}

Fig.~\ref{fig:fisher} quantifies the different degeneracy directions associated with the WN and PS correlation spectra. For the WN model, $\gamma$ has the largest absolute Fisher correlation coefficient with $\rho$ because a shift in $\gamma$ can partially compensate for the mismatch induced by the white-noise model by modifying the SGWB slope across the analysis band. For the PS model, by contrast, $C_{\rm acc}^{\rm PS}(f)$ has exactly the ACC-noise spectral shape, so varying $\rho$ primarily changes the effective normalization of the ACC sector, leading to the strongest degeneracy with $s_{\rm acc}$. These dominant PCCs agree with the leading posterior displacements seen in the violin plots, demonstrating that the inference bias depends not only on the correlation amplitude but also on the frequency dependence of the cross spectrum.

\section{Conclusion and Discussion} \label{sec:conclusion}
We investigate how correlated ACC noise between adjacent test masses can affect SGWB inference with space-based interferometers. Using a moving-orbit forward model, we introduce two phenomenological prescriptions for the single-link ACC cross spectrum and consistently propagate the resulting covariance into the $AET$ interferometric channels. The mock data are analyzed with the $A$ and $E$ autospectra using two Bayesian templates: Corr. Template includes the injected cross-spectral contribution, whereas the Non-Corr. Template omits it. By keeping the SGWB and other instrumental noise spectra, signal response, and likelihood convention fixed, this setup isolates the effect of an incomplete covariance model.

Our main result is that neglecting a nonzero ACC correlation can produce systematic parameter biases even when the signal and instrumental auto-noise models are otherwise correctly specified. In the Non-Corr. analysis, the recovered parameters increasingly depart from their injected values as the correlation amplitude $\rho$ increases, with shifts appearing in both the SGWB and instrumental-noise parameters. In contrast, the Corr. Template remains consistent with the injected parameters over the correlation scan. The overall trend demonstrates that the omitted covariance is sufficient to drive systematic inference shifts.

The direction of the bias is determined by the frequency dependence of the correlation spectrum, rather than by its amplitude alone. For the WN model, whose cross-spectra are distinct from the ACC auto-spectra, the dominant local degeneracy is between $\rho$ and the SGWB spectral slope $\gamma$. For the PS model, the dominant degeneracy instead involves the amplitude of the ACC-noise $s_{\rm acc}$. The corresponding Fisher-matrix directions agree with the leading posterior displacements, showing that the incomplete model preferentially absorbs the missing covariance through the parameter whose spectral effect most closely resembles the omitted contribution. Thus, the correlation coefficient $\rho$ alone is insufficient to characterize the impact of correlated noise on parameter inference. Our results show that a covariance model that captures the relevant instrumental correlations will therefore be an important component of mission-realistic SGWB analyses, particularly as space-based gravitational-wave studies move from idealized sensitivity forecasts toward end-to-end mock-data challenges and inference pipelines.

The present study is deliberately phenomenological and therefore its quantitative conclusions should be interpreted within the adopted assumptions. In particular, the WN and PS cross spectra are not derived from first-principles subsystem models, the extension to $\rho=1$ is diagnostic rather than a realistic in-flight expectation, and we have adopted equal link-level ACC and OMS auto spectra, a single power-law SGWB, and a diagonal $AE$ likelihood. These simplifications isolate the effect of the omitted correlation, but limit the generality of the quantitative conclusions. Future work should incorporate more realistic, frequency-dependent cross spectra and unequal noise levels into the full $AET$ covariance, enabling a systematic study of how correlation-induced biases depend on realistic noise environments, injected signal and noise spectra, and signal-processing choices.

\begin{acknowledgments}
This work is partly supported by the National Key Research and Development Program of China (Grant No.~2021YFC2201901), and the Fundamental Research Funds for the Central Universities.
\end{acknowledgments}


\appendix
\section{Relative Cross-Spectral Power in the TDI Channels} \label{app:tdi_cross_power}

This appendix presents the diagnostic that motivates the channel selection in Section~\ref{sec:TDIobservable}. For each pair of TDI channels $U,V\in\{A,E,T\}$, we normalize the cross-spectrum magnitude by the geometric mean of the corresponding auto spectra
\begin{equation}
    \frac{\abs{S_{UV}(f,t_m)}}{\sqrt{S_U(f,t_m)S_V(f,t_m)}}.
\end{equation}
This dimensionless ratio measures the off-diagonal covariance relative to the characteristic power in the two channels and is therefore more informative than $\abs{S_{UV}}$ alone when the auto spectra have different amplitudes. In the equal-arm limit, the $A$, $E$, and $T$ combinations are orthogonal and the ratio is strongly suppressed. The moving constellation breaks this exact cancellation through unequal and time-dependent delays, so the magnitude and covariance of the three channel pairs must be checked using the same segmented response adopted in the likelihood.

\begin{figure}[t]
    \centering
    \includegraphics[width=0.82\textwidth]{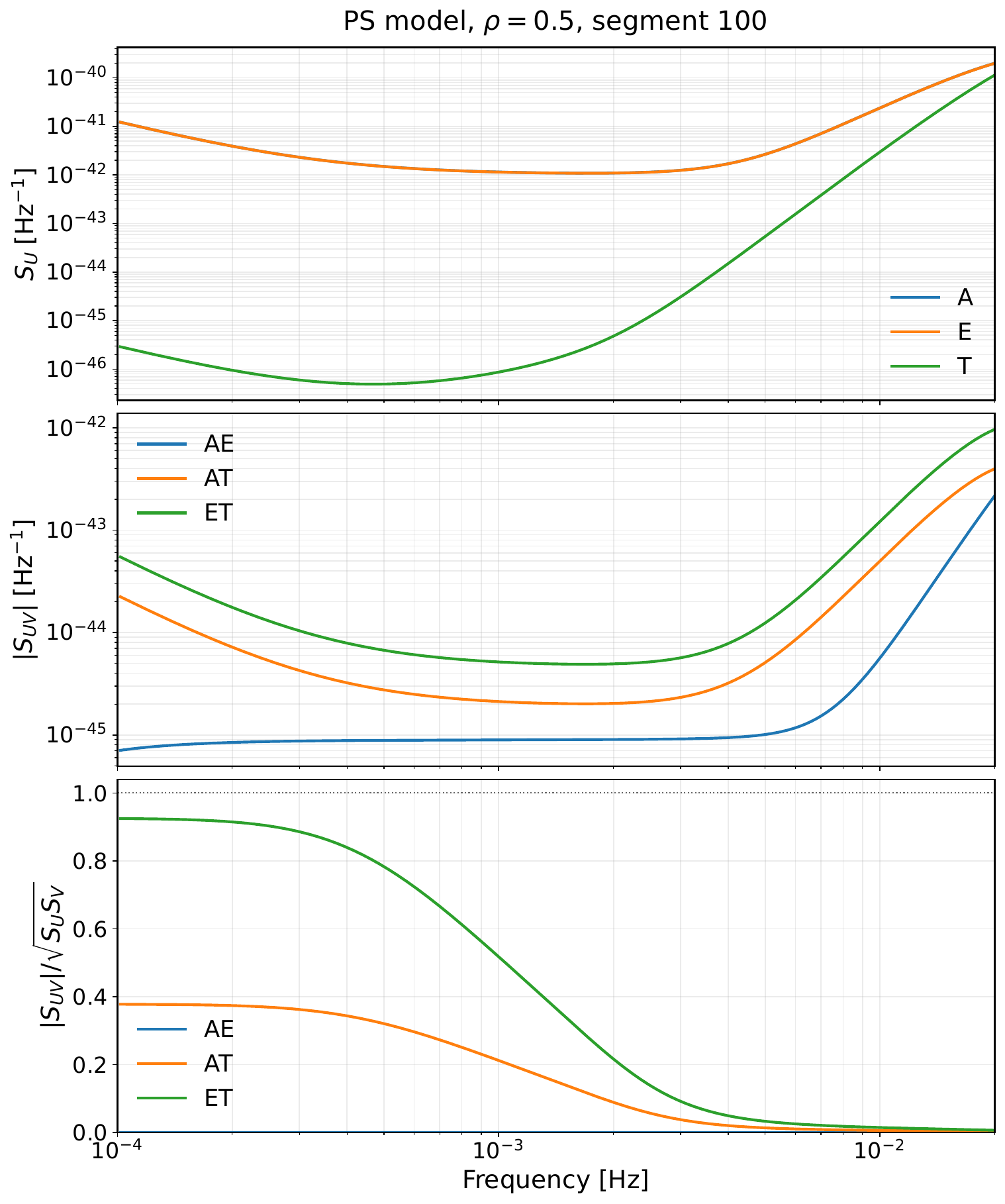}
    \caption{Representative TDI cross-power diagnostic for the PS model with $\rho=0.5$ in STFT segment 100. In the top panel, blue, orange, and green solid lines denote the $A$, $E$, and $T$ auto spectra. In the middle and bottom panels, the same colors denote the $AE$, $AT$, and $ET$ channel pairs, respectively. The middle panel shows $\abs{S_{UV}}$, and the bottom panel shows $\abs{S_{UV}}/\sqrt{S_U S_V}$.}
    \label{fig:appendix_tdi_cross_power}
\end{figure}

Fig.~\ref{fig:appendix_tdi_cross_power} shows a representative example for STFT segment 100 of the PS injection with $\rho=0.5$. The upper panel establishes the relative levels of the $A$, $E$, and $T$ auto spectra, while the middle and lower panels show the absolute and normalized cross powers. The normalized $AT$ and $ET$ cross spectra become appreciable over part of the analysis band, whereas $AE$ remains suppressed. Repeating the same check for the other segments and injected cases yields the same qualitative hierarchy, although the detailed frequency dependence changes with the orbital configuration and noise prescription.
A full $AET$ likelihood therefore cannot consistently retain three auto spectra while setting every off-diagonal term to zero. It must propagate and evaluate $S_{AT}$ and $S_{ET}$, together with the auto spectra, at every frequency and time segment. That calculation is substantially more expensive inside nested sampling and is outside the scope of the present controlled comparison. The small $AE$ ratio, by contrast, supports the diagonal approximation after the data vector is restricted to the $AE$ subspace.

\section{Mock-Data Generation with Correlated Acceleration Noise} \label{app:cacc_analysis}

We generate the correlated ACC noise directly at the single-link level. At each positive-frequency bin, an adjacent pair is represented by $\boldsymbol{\tilde{N}}^{\text{acc}}_{ij,ik}(f)=\{\tilde{N}^{\text{acc}}_{ij}(f),\tilde{N}^{\text{acc}}_{ik}(f)\}^\dagger$ and is drawn from a zero-mean proper complex Gaussian distribution with covariance matrix $\mathbf{M}_{\text{acc}}(f)$ in Eq.~\eqref{eq:accm}. Up to an additive normalization constant, the corresponding log probability is 
\begin{equation}
    \ln p\left[\boldsymbol{\tilde{N}}^{\text{acc}}_{ij,ik}(f)\right]
    =-\left[\boldsymbol{\tilde{N}}^{\dagger\text{acc}}_{ij,ik}(f)\mathbf{M}_{\text{acc}}^{-1}(f)\boldsymbol{\tilde{N}}^{\text{acc}}_{ij,ik}(f)+\ln \abs{\mathbf{M}_{\text{acc}}(f)}\right].
\end{equation}
Joint sampling imposes the prescribed auto spectra and CSD simultaneously. After all positive-frequency bins have been sampled, Hermitian symmetry is imposed, the frequency-domain realizations are transformed to the time domain, and the resulting single-link noises are propagated through the moving-orbit TDI simulation. The resulting TDI observables constitute the mock data analyzed in Section~\ref{sec:PE}.

The theoretical CSD specified by the covariance model should be distinguished from the estimator obtained from a finite stochastic realization. The likelihood contains the ensemble quantity $C_{\rm acc}(f)$, whereas the mock data contain a realization-dependent estimator $\hat{C}_{\rm acc}(f)$. To show explicitly how these two quantities are related after the TDI transformation, consider the auto-spectrum estimator for $U\in\{A,E\}$ 
\begin{equation}
    \hat{S}_{UU}(f,t_m)=\frac{2}{T_c}\tilde{U}(f,t_m)\tilde{U}^{*}(f,t_m).
\end{equation}
Substituting the linear TDI combinations in Eq.~\eqref{eq:TDI_to_siglink} and separating the diagonal and off-diagonal link pairs gives
\begin{equation}
    \begin{split}
        \hat{S}_{UU}(f,t_m)
        &=
        \sum_{ij,lm}[\vb{w}_{U}(f,t_m)]_{ij}[\vb{w}^{*}_{U}(f,t_m)]_{lm}\hat{S}^{\eta}_{ij,lm}(f,t_m) \\
        &=
        \sum_{ij}\left|[\vb{w}_{U}(f,t_m)]_{ij}\right|^2\hat{S}^{\eta}_{ij,ij}(f,t_m) \\
        &+\sum_{ij\neq lm}\,\Re \left\{[\vb{w}_{U}(f,t_m)]_{ij}[\vb{w}^{*}_{U}(f,t_m)]_{lm}\hat{S}^{\eta}_{ij,lm}(f,t_m)\right\}.
    \end{split}
\end{equation}
The coefficients $\vb{w}_U$ act on the link observables $\tilde{\eta}_{ij}$, whereas $C_{\rm acc}$ describes correlations between the original ACC-noise streams. Substituting the ACC terms of Eq.~\eqref{eq:total_data_link} and collecting the contributions from each original noise stream gives
\begin{equation}
    \begin{aligned}
        \tilde{U}^{\rm acc}(f,t_m)
        &=\sum_{ij}b_{U,ij}(f,t_m)\tilde{N}^{\rm acc}_{ij}(f,t_m), \\
        b_{U,ij}(f,t_m)
        &=[\vb{w}_U(f,t_m)]_{ij}
        +[\vb{w}_U(f,t_m)]_{ji}\tilde{D}_{ji}(f,t_m).
    \end{aligned}
    \label{eq:acc_effective_tdi_coefficient}
\end{equation}
For an adjacent ACC pair $\{ij,ik\}$, we define $Q_U^{(ij,ik)}=b_{U,ij}b_{U,ik}^{*}$, which collects the relevant TDI coefficients and delay factors for this correlated pair.
Its contribution to the channel auto spectrum is then written as
\begin{equation}
    \hat{S}^{\rm corr}_{UU}(f,t_m)
    =\Re\left[Q_U^{(ij,ik)}(f,t_m)\hat{C}^{\rm acc}_{ij,ik}(f,t_m)\right].
\end{equation}
In general, both the real and imaginary parts of the estimated CSD can contribute. Separating the real and imaginary parts gives
\begin{equation}
        \hat{S}^{\rm corr}_{UU}=2\left\{\Re[Q_U^{(ij,ik)}]\Re[\hat{C}^{\rm acc}_{ij,ik}]-\Im[Q_U^{(ij,ik)}]\Im[\hat{C}^{\rm acc}_{ij,ik}]\right\}.
\label{eq:cor_term}
\end{equation}

\begin{figure}[t]
    \centering
    \includegraphics[width=0.72\textwidth]{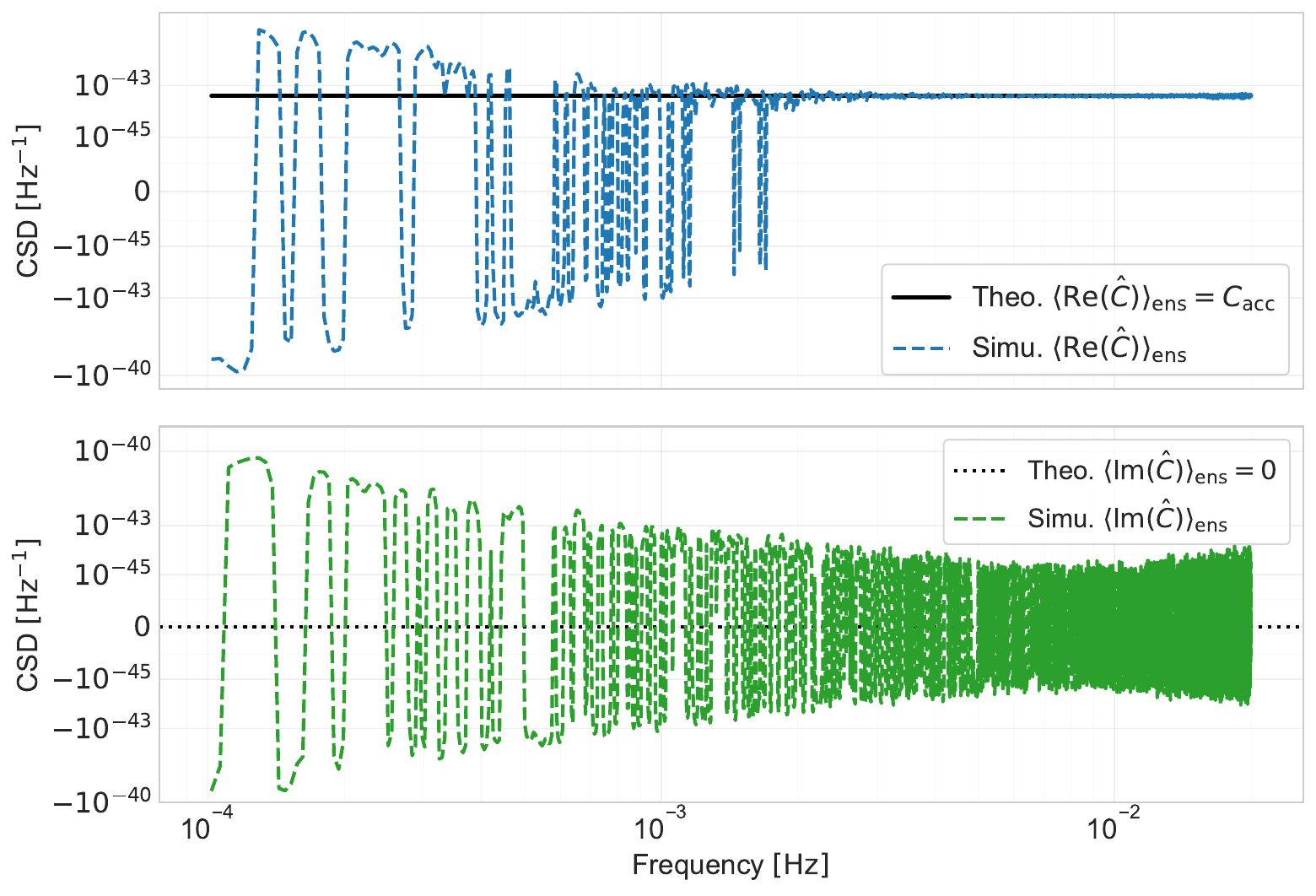}
    \caption{Ensemble validation of the single-link ACC CSD for the WN model with $\rho=0.5$. In the top panel, the black solid line is the injected $C_{\rm acc}(f)$ and the blue dashed line is the ensemble mean of $\Re[\hat{C}]$. In the bottom panel, the black dotted line marks the theoretical value zero and the green dashed line is the ensemble mean of $\Im[\hat{C}]$.}
    \label{fig:appendix_cacc_ensemble}
\end{figure}

The consistency with the theoretical covariance follows after ensemble averaging, since
\begin{equation}
    \expval{\Re \left[\hat{C}^{\text{acc}}_{ij,ik}(f)\right]}_{\text{ens}}=C_{\text{acc}}(f),
    \qquad
    \expval{\Im\left[\hat{C}^{\text{acc}}_{ij,ik}(f)\right]}_{\text{ens}}=0.
\end{equation}
We verify this relation numerically for the WN model with $\rho=0.5$. The recovered CSDs are averaged over ten independent realizations. As shown in Fig.~\ref{fig:appendix_cacc_ensemble}, the mean real component follows the injected $C_{\rm acc}^{\rm WN}(f)$, while the mean imaginary component remains consistent with zero, confirming the expected ensemble behavior.
Taking the ensemble expectation of Eq.~\eqref{eq:cor_term} therefore gives
\begin{equation}
\expval{\hat{S}^{\rm corr}_{UU}(f,t_m)}_{\rm ens}=\text{Re}\left[Q_U^{(ij,ik)}(f,t_m)C_{\rm acc}(f)\right].
\label{eq}
\end{equation}
This is exactly the correlated-noise contribution obtained by propagating the theoretical single-link covariance through the same TDI response. The mock-data and likelihood contributions therefore agree in expectation.

\bibliographystyle{apsrev}
\bibliography{cortm}

\end{document}